\documentclass[12pt]{article}
\usepackage{derivative}
\usepackage{color}
\usepackage{latexsym}
\usepackage{amsmath}
\usepackage{amsfonts}
\usepackage{amssymb}
\usepackage{indentfirst} 
\usepackage{tikz}
\usepackage{enumerate}
\usepackage{authblk}
\usepackage{enumitem}  

\title{The families of Hamiltonians sharing common symmetry structure }
\author[1]{C. Gonera \thanks{e-mail: cezary.gonera@uni.lodz.pl}}
\author[1]{J. Gonera \thanks{e-mail: joanna.gonera@uni.lodz.pl}}
\author[1]{A. Jasiński }
\author[1]{P. Kosiński \thanks{e-mail: piotr.kosinski@uni.lodz.pl}}

\affil[1]{\small Faculty of Physics and Applied Informatics,
University of Lodz, Pomorska 149/153, 90-236 Łódź, Poland.}

\date{\today}
\begin{document}
\maketitle

\begin{abstract}
We describe a general procedure which allows to construct, starting from a given Hamiltonian, the whole family of new ones sharing the same set of unparameterized trajectories in phase space. The symmetry structure of this family can be completely characterized provided the symmetries of initial Hamiltonian are known. Our approach covers numerous models considered in literature as well as it allows to construct novel ones. It provides a far reaching generalization of  Hietarinta et al. coupling-constant metamorphosis method and another proof of Darboux theorem.

\end{abstract}

\section{Introduction}
\par

Hamiltonian dynamical systems play an important role both in physics and mathematics. In the literature one encounters many examples which are interesting due to their relevance to physics and/or intriguing mathematical properties. It happens quite often that the systems defined by different Hamiltonians share the same unparameterized trajectories in phase space. They are, however, related in rather nontrivial way since the relevant time reparameterization varies, in general, from trajectory to trajectory, i.e. is a function of canonical variables. The Hamiltonians sharing unparameterized phase space trajectories admit the same set of integrals of motion which do not depend explicitly on time. This implies that they are to the same extent integrable. In particular, they are (or not) simultaneously (super)integrable.
The classical example of such systems is provided by Darboux theorem \cite{dar}, \cite{alb}. Other very interesting examples have been constructed by the related method proposed by Hietarinta et.al. \cite{hit} under the name of coupling-constant metamorphosis. It allowed, for example, to relate the Henon-Heiles \cite{h-h} and Holt \cite{holt1, holt} Hamiltonians.

In the present paper we propose a fairly general method of constructing, starting from a given Hamiltonian, a whole family of related Hamiltonians sharing the same set of unparameterized phase space trajectories. It provides also an explicit relation between the Poisson algebras of integrals of motion. One can say that our approach yields a complete description of the symmetry structure of the resulting Hamiltonians provided the symmetry of the initial one is known. In general, the final Hamiltonian doesn't have the standard form even though we start with the standard one (i.e. a sum of a non-degenerate quadratic form in momenta plus a function of generalized coordinates only). However, in many cases one obtains the Hamiltonian corresponding to the metric in configuration space conformally equivalent to that of the initial Hamiltonian. The method proposed generalizes the Darboux and Hietarinta et al. approaches as well as it allows to describe and generalize many examples of dynamical systems encountered in the literature. In particular it provides an interesting relationship between nonrelativistic and relativistic dynamics.

The paper is organized as follows. In Sec. II the general formalism is presented while Sec. III is devoted to some examples discussed recently in the literature \cite{1,2,3,11,12,15,16,18}. We consider oscillator- and Kepler-related systems as well as Henon-Heiles-like systems on curved configuration space which, as far as we can see, haven't been discussed previously (see, however \cite{bal}). The relativistic-nonrelativistic correspondence is described in Sec. IV. Darboux  and Hietarinta   approaches are considered in Sec. V. Finally, Sec. VI is devoted to some conclusions.

\section{General scheme}
\par
Assume we have a Hamiltonian system defined in 2N-dimensional phase space. The relevant Hamiltonian reads:

\begin{equation}
\label{r2.1}
H=H({\bf q}, {\bf p}, {\boldsymbol{\lambda}})
\end{equation}
\,
\\
where ${\bf q}\equiv(q_1,...,q_N)$, ${\bf p}\equiv (p_1,...,p_N)$ are canonical variables while ${\boldsymbol\lambda} \equiv (\lambda_1,...,\lambda_M)$ - the set of parameters (masses, coupling constants, frequencies etc.) characterizing our system. One defines the new Hamiltonian $\tilde H ({\bf{q}, {p}})$ as follows. Let $\tilde H$ be a new variable; replace the parameters $\lambda_{\alpha}$ by the functions

\begin{equation}
\label{r2.2}
\lambda_{\alpha}= \lambda_{\alpha}(\tilde H).
\end{equation} 
\,
\\
Assume that they are chosen in such a way  that the equation

\begin{equation}
\label{r2.3}
\tilde{H}({\bf{q}, p}) = H({\bf{q}, p}, {\boldsymbol{\lambda}}(\tilde H({\bf{q}, p}))
\end{equation}
\,
\\
has the unique solution $\tilde H(\bf{q}, \bf{p})$. In particular,

\begin{equation}
\label{r2.4}
\Omega ({\bf q,  p}) \equiv 1- \frac{\partial H}{\partial \lambda_{\alpha}}\frac{\partial \lambda_{\alpha}}{\partial \tilde H}\neq 0
\end{equation}
\,
\\
$\tilde{H}({\bf q,  p})$ defines new dynamics on our phase space. Obviously, we may assume that the functions $\lambda_{\alpha}$, eq. (\ref{r2.2}), still depend on some parameters, $\lambda_{\alpha}=\lambda_{\alpha}(\tilde H, \mu)$. Both Hamiltonian dynamics, generated by $H$ and $\tilde H$, are closely related. One finds:

\begin{equation}
\label{r2.5}
\begin{split}
& \frac{\partial H}{\partial q_i} = \Omega({\bf {q}, p}) \frac{\partial \tilde H}{\partial q_i}\\
& \frac{\partial H}{\partial p_i} = \Omega({\bf {q}, p}) \frac{\partial \tilde H}{\partial p_i}
\end{split}
\end{equation}
\,
\\
Therefore,

\begin{equation}
\label{r2.6}
\begin{split}
& \odv{q_i}{t}=\pdv{H}{p_i} =\Omega({\bf q,p})\pdv{\tilde H}{p_i}=\Omega({\bf q, p}) \odv{q_i}{\tilde t}\\
& \odv{p_i}{t}=-\pdv{H}{q_i} =-\Omega({\bf q,p})\pdv{\tilde H}{q_i}=\Omega({\bf q, p}) \odv{p_i}{\tilde t};
\end{split}
\end{equation}
\,
\\
here $t\left( \tilde{t} \right)$ is the time parameter for the dynamics generated by $H(\tilde H)$. Let ${\bf q= \tilde q}\left(\tilde{t} \right)$, ${\bf p= \tilde p}\left(\tilde{t}\right)$ be any particular solution to the Hamiltonian equations for $\tilde H$, corresponding to some initial data (in particular, energy $\tilde E$). Then $\Omega$ becomes a function of time $\tilde t$ only, $\Omega=\Omega(\tilde t)$. Let us relate $t$ to $\tilde t$ by the equation

\begin{equation}
\label{r2.7}
\odv{\tilde t}{t} = \Omega.
\end{equation}
\,
\\
By comparing eqs.  (\ref{r2.6}) and  (\ref{r2.7}) we conclude that

\begin{equation}
\label{r2.8}
\begin{split}
&q_i=\tilde q_i(\tilde t(t))\\
&p_i=\tilde p_i(\tilde t(t)),  \quad \quad i=1,2,...,N
\end{split}
\end{equation}
\,
\\
are the solutions to the Hamiltonian equations for the Hamiltonian

\begin{equation}
\label{r2.9}
H=H({\bf q,p},{\boldsymbol \lambda}(\tilde E))
\end{equation}
\,
\\
Obviously, this works in the reverse direction as well.

It follows then that the set of all trajectories in phase space generated by $\tilde H({\bf q, p})$ coincides with that for $H({\bf q,p},{\boldsymbol \lambda}(\tilde E))$; they differ only by time parameterization (this difference varies from trajectory to trajectory). Therefore, the (super)integrability of $H$ implies the (super)integrability of $\tilde H$ and reverse.\\

One can easily relate the algebras of integrals of motion for both dynamical systems. For any function on phase space,

\begin{equation}
\label{r2.10}
F\equiv F({\bf q,p};{\boldsymbol \lambda})
\end{equation}
\,\\
define

\begin{equation}
\label{r2.11}
\tilde F\equiv F({\bf q,p};{\boldsymbol \lambda}(\tilde{H}({\bf q,p}))).
\end{equation}
\,
\\
Then
\begin{equation}
\label{r2.12}
\begin{split}
& \pdv{\tilde F}{q_i} = \pdv{F}{q_i} + \pdv{F}{\lambda_\alpha} \pdv{\lambda_\alpha}{\tilde H} \pdv{\tilde H}{q_i}\\
& \pdv{\tilde F}{p_i} = \pdv{F}{p_i} + \pdv{F}{\lambda_\alpha} \pdv{\lambda_\alpha}{\tilde H} \pdv{\tilde H}{p_i}.
\end{split}
\end{equation}
\,
\\
In particular, let

\begin{equation}
\label{r2.13}
C=C({\bf q,p},{\boldsymbol \lambda})
\end{equation}
\,\\
be any integral of motion of the dynamics generated by H. Then one easily finds

\begin{equation}
\label{r2.14}
\{ \tilde C, \tilde H \} =0
\end{equation}
\,\\
implying that $\tilde C$ is also an integral of motion; this follows also directly from the coincidence of trajectories for both Hamiltonians.

Moreover, for any two integrals of motion we have

\begin{equation}
\label{r2.15}
\{ \tilde C_a, \tilde C_b \} = \widetilde{\{C_a,C_b\}}
\end{equation}
\,\\
If $\{ C_a\}$ is the set of all functionally independent integrals of motion for $H$, then

\begin{equation}
\label{r2.16}
\{ C_a, C_b \} = F_{ab}({\bf C},{\boldsymbol \lambda}) 
\end{equation}
\,\\
with $F_{a,b}$ being some functions. Equations (\ref{r2.15}) implies

\begin{equation}
\label{r2.17}
\{ \tilde C_a, \tilde C_b \} = F_{ab}({\bf \tilde C}, \tilde{\boldsymbol{\lambda}}(\tilde H))
\end{equation}
\,
\\
Concluding, the algebras of symmetries for both systems are closely related.

Finally, let us consider a particular form of the transformation discussed above. We start with the Hamiltonian (\ref{r2.1}), assuming, for simplicity (but without loosing generality) that there is one parameter $\lambda$. Obviously, nothing will change if we start with the Hamiltonian

\begin{equation}
\label{d1}
H'=H({\bf q,p},\lambda) + d
\end{equation}
\,\\
with $d$ being an arbitrary constant. We make again the replacement $\lambda 
\rightarrow \lambda (\tilde{H})$ together with $d \rightarrow \tilde{H}-d$ one. Then eq.  (\ref{r2.3}) takes the form

\begin{equation}
\label{d2}
0=H({\bf{q,p}},\lambda(\tilde{H})) - d
\end{equation}
\,\\

Assuming that it can be solved with respect to $\tilde{H}$ one finds

\begin{equation}
\label{d3}
\tilde{H} = \tilde{H} ({\bf{q,p}},d)
\end{equation}
\,\\
Obviously, our previous reasoning is still valid including that concerning integrals of motion and their algebra. We conclude that the trajectories described by the initial Hamiltonian $H$ with the coupling $\lambda(\tilde{E})$,  corresponding to the energy $E=d$ coincide (up to the reparameterization) with those described by the Hamiltonian $\tilde{H}$ with
the coupling $d=E$ and energy $\tilde{E}$. This is the generalized version of the so called coupling-constant metamorphosis method (see Sec. IV).

\section{Examples}
\subsection{Oscillator-related systems}
\par
Consider the three-dimensional isotropic harmonic oscillator

\begin{equation}
\label{r3.17}
H=\frac{1}{2} \left( \vec{p}\,^2 + \omega ^2 \vec{q}\,^2 \right).
\end{equation}
It is superintegrable. To reveal the structure of the algebra of integrals of motion define the classical counterparts of creation/annihilation operators:

\begin{equation}
\label{r3.18}
a_i \equiv \frac{1}{\sqrt{2\omega}} \left(p_i - \imath \omega q_i \right), \qquad i=1,2,3
\end{equation}
\,\\
Then

\begin{equation}
\label{r3.19}
\{ a_i, \bar a_j\} = -\imath \delta_{ij}.
\end{equation}
\,\\
Further, let $\Lambda_a$ , $a=1,2,...,8$ be the Gell-Mann matrices obeying $SU(3)$ Lie algebra commutation rules

\begin{equation}
\label{r3.20}
[\Lambda_a,\Lambda_b] = \imath {d_{ab}}^c \Lambda _c
\end{equation}
\,\\
Then the functions

\begin{equation}
\label{r3.21}
C_a \equiv \frac{1}{2} a_i(\Lambda_a)_{ij}\overline{a}_j
\end{equation}
\,\\
are real integrals of motion obeying

\begin{equation}
\label{r3.22}
\{C_a,C_b\}={d_{ab}}^c C_c.
\end{equation}
\,\\
Let us now modify the parameter $\omega ^2$:

\begin{equation}
\label{r3.23}
\omega ^2 (\tilde H) \equiv \omega ^2 - 2\mu \tilde H, \qquad \mu \in R
\end{equation}
\,\\
Following the prescription of the previous section we find the Hamiltonian \cite{1,2}

\begin{equation}
\label{r3.24}
\tilde H = \frac{1}{2} \frac{\vec{p}\,^2 +\omega^2 \vec{q}\,^2}{1 + \mu \vec q \,^2}
\end{equation}
\,\\
For $\mu \geq 0$ $\tilde H$ is non-singular everywhere; for $\mu < 0$ some care must be exercised. The function $\Omega ({\bf q,p})$, eq.  (\ref{r2.4}), takes the form

\begin{equation}
\label{r3.25}
\Omega ({\bf q,p})= 1 + \mu {\vec{q}}\,^2
\end{equation}
\,\\
For simplicity consider the case $\mu \geq 0$. Assuming $\tilde E < \displaystyle{\frac{\omega ^2}{2\mu}}$ we have $\omega^2(\tilde H) > 0$ (bounded motion). Eqs.  (\ref{r3.18}), (\ref{r3.20}) are replaced by

\begin{equation}
\label{r3.26}
\tilde{a}_i = \frac{1}{\sqrt{2\omega (\tilde H)}} \left( p_i - \imath \omega(\tilde H)q_i\right), \quad i=1,2,3
\end{equation}

\begin{equation}
\label{r3.27}
\tilde{C}_a =\frac{1}{2}\tilde{a}_i\left(\Lambda_a\right)_{ij} \overline{\tilde a}_j
\end{equation}
\,\\
According to the discussion of Sec. II $\tilde C_a$ are integrals of motion obeying $SU(3)$ algebra 

\begin{equation}
\label{r3.28}
\{\tilde C_a,\tilde C_b\}={d_{ab}}^c \tilde C_c.
\end{equation}
\,\\
Let us note that eq. (\ref{r3.19}) is no longer valid.

For $\tilde E = \displaystyle{\frac{\omega ^2}{2\mu}}$ ( $\tilde E > \displaystyle{\frac{\omega ^2}{2\mu}}$) the relevant symmetry algebra gets modified appropriately.

Let us now add the linear term to harmonic potential,

\begin{equation}
\label{r3.29}
H=\frac{1}{2} \left( \vec{p}\,^2 + \omega ^2 \vec{q}\,^2 \right) + \vec{k}\cdot\vec{q}
\end{equation}
\,\\
with $\vec{k}$ being a constant vector.\\
By making the canonical transformation 

\begin{equation}
\label{r3.30}
\begin{split}
&\vec{Q} = \vec{q} +\frac{\vec{k}}{\omega^2}\\
&\vec{P} =\vec{p}
\end{split}
\end{equation}
\,\\
$H$ reduces to harmonic oscillator:
\begin{equation}
\label{r3.31}
H=\frac{1}{2} \left( \vec{P}\,^2 + \omega ^2 \vec{Q}\,^2 \right) + \text{const}
\end{equation}
\,\\
Therefore, the superintegrability is preserved together with the structure of the symmetry algebra.\\
Making the substitution (\ref{r3.23}) one obtains the Hamiltonian studied in \cite{3}

\begin{equation}
\label{r3.32}
\tilde H = \frac{1}{1+\mu \vec{q}\,^2} \left( \frac{1}{2} \left( \vec{p}\,^2 + \omega ^2 \vec{q}\,^2 \right) + \vec{k}\cdot\vec{q} \right)
\end{equation}
\,\\
Again, assuming $\tilde E < \displaystyle{\frac{\omega ^2}{2\mu}}$ and defining

\begin{equation}
\label{r3.33}
a_i = \frac{1}{\sqrt{2\omega (\tilde H)}} \left( p_i - \imath \omega(\tilde H)q_i - \frac{\imath k_i}{\omega(\tilde H)}\right), \quad i=1,2,3
\end{equation}
\,
\\
we conclude that the integrals $\tilde C_a$, eq.  (\ref{r3.27}), obey $SU(3)$ algebra. The price one has to pay for having $SU(3)$ algebra as the symmetry algebra is that the integrals $\tilde{C}_a$ are not polynomial in momenta. However, when multiplied  by $\omega ^3(\tilde H)$  they become polynomials; the modified integrals of motion obey then the Poisson commutation rules with $\tilde H$-dependent structure constants.

As the next example consider the generalization of Smorodinsky-Winternitz system \cite{4,5,6,7,8,9,10,11,12}. The relevant Hamiltonian reads

\begin{equation}
\label{r3.34}
H=\frac{1}{2}\left( \vec{ p}\,^2 +\sum_{i=1}^3 \omega^2_i q^2_i \right) + \sum_{i=1}^3 \frac{k_i}{q^2_i}
\end{equation}
\,\\
with $k_i>0$, $i=1,2,3$. $H$ is obviously integrable being separable in Cartesian coordinates. It is straightforward to find the action variables $I_k$, $k=1,2,3$; in terms of them the Hamiltonian takes the form

\begin{equation}
\label{r3.35}
H=2\sum_{k=1}^3 \omega_kI_k + \sum_{k=1}^3 \sqrt{2k}\omega_k
\end{equation}
\,
\\
Assuming all ratios $\displaystyle{\frac{\omega_i}{\omega_j}}$ rational we conclude that it is maximally superintegrable \cite{13}. Putting $\omega_k=n_k\omega$, with $n_k$ natural, one can write

\begin{equation}
\label{r3.36}
H=\frac{1}{2}\left(\vec p \,^2 +\omega^2\sum_{k=1}^3n^2_kq^2_k\right) + \sum_{i=1}^3\frac{k_i}{q^2_i}
\end{equation}
\,\\
The integrals of motion can be chosen to obey Lie algebra commutation rules with respect to Poisson brackets \cite{14};  it is again $SU(3)$ algebra.\\
Let us now replace $\omega^2\rightarrow \omega^2 -2\mu \tilde{H}$. Then

\begin{equation}
\label{r3.37}
\tilde{H}=\frac{1}{\left(1+\mu \displaystyle{\sum_{k=1}^3n_k^2q_k^2}\right)}
\left(\frac{1}{2}\vec{p}\,^2 +\frac{\omega^2}{2} \sum_{k=1}^3 n_k^2q^2_k +\sum_{i=1}^3\frac{k_i}{q^2_i}\right)
\end{equation}
\,
\\
For regularity we assume $\mu>0$. According to the discussion of the previous section $\tilde{H}$ is superintegrable. Assuming $\omega^2 -2\mu \tilde{E}>0$ we find also that the integrals of motion can be chosen in such a way as to obtain $SU(3)$ algebra. The particular case of $\tilde{H}$, $n_1=n_2=1$, $n_3=2$ has been considered in \cite{3}.

It is instructive to analyze in more detail the integrals of motion for the Hamiltonian (\ref{r3.36}) (and, consequently, the Hamiltonian  (\ref{r3.37})). The angle variables for the dynamics (\ref{r3.36}) read

\begin{equation}
\label{r3.38}
\begin{split}
\cos{\varphi_i} =\frac{E_i -n^2_i\omega^2q_i^2}{\sqrt{E_i^2 -2k_in_i^2\omega^2}}\\
\sin{\varphi_i} =\frac{E_i -n^2_i\omega^2p_i^2}{\sqrt{E_i^2 -2k_in_i^2\omega^2}}
\end{split}
\end{equation}
\,
\\
with $E_i=2\omega n_iI_i + \sqrt{2k_i}\omega n_i$ being partial energies. It is easy to write out the general form of the integral of motion \cite{18}

\begin{equation}
\label{r3.39}
C=\sum_{\{m\} }F_{\{m\} } ({\bf{I}})\prod_{k=1}^3 \left( e^{\imath\varphi_k}\right)^{m_k}
\end{equation}
\,
\\
with $\{m\}\equiv \{m_1,m_2,m_3\}$ being arbitrary triples of integers obeying

\begin{equation}
\label{r3.40}
m_1n_1 +m_2n_2 +m_3n_3=0
\end{equation}
\,
\\
and $F_{\{m\} } (\bf{I})$ - arbitrary functions. Obviously, only five of these integrals can be chosen as functionally independent. It is also clearly seen from eqs. (\ref{r3.38}),(\ref{r3.39}) that one can choose the integrals as polynomials in momenta. For example, for $n_1=n_2=1$, $n_3=2$, the quadratic integral $K_4$ in \cite{3} may be written as

\begin{equation}
\label{r3.41}
\omega^2 K_4 = E_1 \cdot E_2 - \sqrt{E^2_1 -2k_1\omega^2}\sqrt{E^2_2 -2k_2\omega^2}\cos{(\varphi_1-\varphi_2)}
\end{equation}
\,
\\
Now, for $\omega^2 -2\mu \tilde{E}>0$ the relevant integrals $\tilde{C}$ for Hamiltonian $\tilde{H}$ can be obtained along the lines sketched in previous section.\\
Let us come back to the Hamiltonian (\ref{r3.29}). Adding a constant $h$ to the Hamiltonian and making the substitutions $\omega^2 \rightarrow \omega^2 -2\mu \tilde{H}$, $\vec k \rightarrow\vec k -\tilde{H}\vec l$, $h \rightarrow h-h_0\tilde{H}$ we find

\begin{equation}
\label{r3.51}
\tilde{H}=\frac{1}{\mu \vec q\,^2 + \vec l \cdot \vec q + h_0 +1}\left( \frac{\vec p\, ^2}{2} + \frac{\omega^2}{2} \vec q \,^2 + \vec k \cdot \vec q + h \right)
\end{equation}
\,
\\
Considering two dimensions and putting $l_i=\delta_{i1}$, $k_i=\delta_{i2}$, $\mu=0$, $h_0=-1$ we arrive at the one of the Hamiltonians considered in \cite{18}  (eq. (64) therein).\\
Further, let us start with the two-dimensional Hamiltonian

\begin{equation}
\label{r3.52}
H=\frac{1}{2}\vec p \,^2 + \frac{\omega^2}{2}\left(4q_1^2 + q_2^2 \right) +k_1q_1 + \frac{k_2}{q_2^2} +h
\end{equation}
\,
\\
which is superintegrable. We put $h \rightarrow h_0+\tilde{H}$, $k_1 \rightarrow -\tilde{H}$; then

\begin{equation}
\label{r3.53}
\tilde H = \frac{\vec p \,^2}{2q_1} +\frac{\omega^2 (4q_1^2 + q_2^2)}{2q_1} + \frac{h_0}{q_1} +\frac{k_2}{q_1q_2^2}
\end{equation}
\,
\\
which is the Hamiltonian (59) from \cite{18}.

\subsection{Kepler-related systems}

As a next example consider Kepler-related Hamiltonians. The initial Hamiltonian for the textbook Kepler problem reads

\begin{equation}
\label{r3.42}
H=\frac{1}{2} \vec p \,^2 +\frac{\alpha}{| \vec q \,^2 | }, \quad \alpha<0
\end{equation}
\,
\\
We make the replacement $\alpha \rightarrow \alpha + \beta
 \tilde{H}$ which yields
 
\begin{equation}
\label{r3.43}
\tilde{H}=\left(\frac{|\vec{q}\,|}{|\vec{q}\,|-\beta}\right) \left(\frac{1}{2} \vec p\, ^2 + \frac{\alpha}{|\vec{q}\,|}\right)
\end{equation}
\,\\
For regularity we assume $\beta <0$. $\tilde{H}$ is superintegrable. Except energy and angular momentum it admits further integrals of motion in the form of generalized Runge-Lenz vector
 
\begin{equation}
\label{r3.44}
\tilde{\vec A} = \vec A - \frac{m\beta\tilde{H}\vec q}{|\vec q\,|^2}
\end{equation}
\,
\\
where $\vec A$ is a standard Runge-Lenz vector.\\

The Kepler dynamics can be generalized in a way still preserving superintegrability \cite{11},  \cite{12},  \cite{15}:

\begin{equation}
\label{r3.45}
H=\frac{1}{2}\vec p\,^2 + \frac{\alpha}{|\vec{q}\,|} + \sum_{i=1}^3 \frac{k_i}{q^2_i}
\end{equation}
\,
\\
Under the same substitution $\alpha \rightarrow \alpha + \beta
 \tilde{H}$, one finds
 
\begin{equation}
\label{r3.46}
\tilde{H}=\left(\frac{|\vec{q}\,|}{|\vec{q}\,|-\beta}\right) \left(\frac{1}{2} \vec p\, ^2 + \frac{\alpha}{|\vec{q}\,|} + \sum_{i=1}^3 \frac{k_i}{q^2_i} \right)
\end{equation}
\,
\\
It defines, according to our findings, the superintegrable dynamics. For a more detailed discussion of its properties we refer the reader to \cite{3}. The integrals of motion described there can be derived using the technique introduced in the present paper.

\subsection{Curved configuration space}
Both harmonic oscillator and Kepler system with or without additional term $\displaystyle{ \sum_{i=1}^3 \frac{k_i}{q^2_i}}$ (eqs. (\ref{r3.36}) and (\ref{r3.45})) can be generalized to the case of configuration space of constant curvature \cite{16}. Our procedure works also in this case. For example, the curvature-dependent version of Smorodinsky-Winternitz system leads to the following Hamiltonian

\begin{equation}
\label{r3.47}
\tilde{H}=\frac{1}{1-\mu T^2_\kappa (r)}\left(\frac{1}{2} \left(p_r^2 + \frac{1}{s^2_\kappa (r)}\left(p^2_\theta +\frac{p^2_\varphi}{\sin^2(\theta)}\right) \right) +\frac{1}{2} \alpha^2 T^2_\kappa (r) + \sum_{i=1}^3 \frac{k_i}{q^2_{\kappa_i}} \right)
\end{equation}
\,
\\
where

\begin{equation}
\label{r3.48}
s_\kappa (r)=\begin{cases}
\frac{1}{\sqrt{\kappa}}\sin(\sqrt{\kappa}r),&\quad \kappa>0\\
r,&\quad \kappa=0\\
\frac{1}{\sqrt{-\kappa}}\sinh(\sqrt{-\kappa}r),&\quad \kappa<0\\
\end{cases}
\end{equation}
\,
\\
\begin{equation}
\label{r3.49}
T_\kappa (r)=\begin{cases}
\frac{1}{\sqrt{\kappa}}\tan(\sqrt{\kappa}r),&\quad \kappa>0\\
r,&\quad \kappa=0\\
\frac{1}{\sqrt{-\kappa}}\tanh(\sqrt{-\kappa}r),&\quad \kappa<0\\
\end{cases}
\end{equation}
\,
\\
and

\begin{equation}
\label{r3.50}
\begin{split}
q_{\kappa_1}&=s_\kappa(r)\sin \theta \cos \varphi\\
q_{\kappa_2}&=s_\kappa(r)\sin \theta \sin \varphi\\
q_{\kappa_3}&=s_\kappa(r)\cos \theta
\end{split} 
\end{equation}
\,
\\
These systems are discussed in detail in \cite{16}. It follows immediately that they are superintegrable and their integrals of motion can be obtained following the procedure described in the present paper.

\subsection{Henon-Heiles related system}
The general Hamiltonian describing the Henon-Heiles system reads \cite{h-h}
\begin{equation}
\label{r3.51a}
H=\frac{1}{2}\left(p^{2}_{x}+p^{2}_{y}\right)+\frac{A}{2}x^{2}+\frac{B}{2}y^{2}+D\left(yx^{2}-\frac{1}{3}\mu y^{3}\right)+C
\end{equation}
\,
\\
In general, $ H $ describes nonintegrable dynamics except three cases \cite{bla}
 \begin{enumerate}[label=(\roman*)]
        \item $ \mu = -6\; ,\; A,B\, $ - arbitrary
        \item $ \mu = -1\; ,\;A=B $
        \item $ \mu = -16\; ,\;B=16A $
    \end{enumerate}
    
\par By applying the general method described above one can define a number of Hamiltonians sharing the same phase space trajectories. Below we present few examples.\\
Consider first the case \textit{(i)}.  Upon making the replacement
\begin{equation}
\label{r3.52a}
A\rightarrow A-2a\tilde{H}\; ,\;\; B\rightarrow B-2b\tilde{H}\; ,\;\; C=0
\end{equation}
\,
\\
one arrives at
\begin{equation}
\label{r3.53a}
\tilde {H}=\frac{\frac{1}{2}\left(p_{x}^{2}+p_{y}^{2}\right)+\frac{A}{2}x^{2}+\frac{B}{2}y^{2}+D\left(x^{2}y+2y^{3}\right)}{1+ax^{2}+by^{2}}
\end{equation}
\,
\\
The additional integral of motion reads
\begin{equation}
\label{r3.54}
\tilde {K}=K-8aD\tilde {H}yx^{2}+\left(\left(2\left( aB+bA \right)-1baA\right)\tilde {H}+4a \left(4a-b\right)\tilde{H}^{2}\right)x^{2}+2\left(b-4a\right)\tilde{H}p^{2}_{x}
\end{equation}
\,
\\
where 
\begin{equation}
\label{r3.55}
K=2\left(4A-B\right)\left(\frac{p^{2}_{x}}{2}+\frac{A}{2}x^{2}\right)+4D\left(p_{x}\left(xp_{y}-yp_{x}\right)
+x^{2}\left(Ay+D(x^{2}+y^{2})\right)\right)
\end{equation}
\,
\\
is the integral of motion for initial dynamics \cite{bla}.\\
On the other hand, keeping first two substitutions (\ref{r3.52a}) intact, we can put $ C=\tilde {H} $ which results in
\begin{equation}
\label{r3.56}
\tilde {H}=\frac{\frac{1}{2}\vec{p}\,^{2}+\frac{A}{2}x^{2}+\frac{B}{2}y^{2}+D\left(x^{2}y+2y^{3}\right)}{ax^{2}+by^{2}}
\end{equation}
\,
\\
Let us recall that, according to the general discussion in Sec. II, the trajectories described by the Hamiltonian $ \tilde{H} $, eq. (\ref{r3.56}), with the energy $ \tilde{E} $, coincide with those obtained from $H$ with the vanishing energy and parameters $A, B$ replaced by $A-2a\tilde{E}$, $B-2b\tilde{E}$, respectively. In other words, the family of trajectories defined by the given Hamiltonian $\tilde{H}$ with arbitrary energy $\tilde{E}$ is mapped to the trajectories of the family of Hamiltonians with parameters $A-2a\tilde{E}$, $B-2b\tilde{E}$ and vanishing energy. \\
The corresponding additional integral of motion is defined again by eq. (\ref{r3.54}) where $\tilde{H}$ is given now by (\ref{r3.56}).

Similar reasoning may be applied to the remaining cases (\textit{ii}), (\textit{iii}). For example, starting with
\begin{equation}
\label{r3.57}
H=\frac{1}{2}\vec{p}\,^{2}+\frac{A}{2}\left(x^{2}+16y^{2}\right)+D\left(x^{2}y+\frac{16}{3}y^{3}\right)+C
\end{equation}
\,
\\
and putting $A\rightarrow A-2a\tilde{H}$, $C=0$ we find
\begin{equation}
\label{r3.58}
\tilde{H}=\frac{\frac{1}{2}\vec{p}\,^{2}+\frac{A}{2}\left(x^{2}+16y^{2}\right)+D\left(x^{2}y+\frac{16}{3}y^{3}\right)}{1+a\left(x^{2}+16y^{2}\right)}\quad \text{,}
\end{equation}
\,
\\
while
\begin{equation}
\label{r3.59}
\tilde{K}=K-12a\tilde{H}x^{2}p^{2}_{x}+8aD\tilde{H}x^{4}y-12a\left(A\tilde{H}+a\tilde{H}^{2}\right)x^{4}
\end{equation}
\,
\\
and []
\begin{equation}
\label{r3.60}
K=3p^{4}_{x}+6Ax^{2}p^{2}_{x}+12Dyx^{2}p^{2}_{x}-4DAx^{4}y-4D^{2}x^{4}y^{2}+3A^{2}x^{2}-\frac{2}{3}D^{2}x^{6}-4Dx^{3}p_{x}p_{y}
\end{equation}
\,
\\
On the other hand, taking $C=\tilde{H}$, instead of $C=0$, one finds
\begin{equation}
\label{r3.61}
\tilde{H}=\frac{\frac{1}{2}\vec{p}\,^{2}+D\left(x^{2}y+\frac{16}{3}y^{3}\right)}{a\left(x^{2}+16y^{2}\right)}
\end{equation}
\,
\\
where $\tilde{K}$ is given by eq. (\ref{r3.59}) with $\tilde{H}$ defined by (\ref{r3.61}).

\section{The relativistic-nonrelativistic correspondence}
\par

An interesting picture emerges if one applies the algorithm described above to relativistic dynamics. Consider the charged relativistic particle moving in a static electromagnetic field. Then one can choose the potentials $\varphi$ and $\vec{A}$ to be time independent. The relevant Hamiltonian reads

\begin{equation}
\label{rel.1}
H=c\sqrt{( \vec{p}-\frac{e}{c}\vec{A})^2 + m^2c^2} + e\varphi + D
\end{equation}
\,
\\
where $\vec{p}$ stands for the canonical momentum while $D$ is an arbitrary constant. Let us make the substitution
\begin{equation}
\label{rel.2}
m^2 \rightarrow m^2(\tilde{H}) \equiv m^2 - \frac{2m \tilde H}{c^2}
\end{equation}
\,
\\
\begin{equation}
\label{rel.3}
D \rightarrow D(\tilde{H}) \equiv \tilde H - D\; \;\text {;}
\end{equation}
\,
\\
in order to keep $m^2(\tilde H)$ nonnegative we have to assume 

\begin{equation}
\label{rel.4}
\tilde{H}\leq \frac{m c^2}{2}
\end{equation}
\,
\\
The general scheme described in previous sections results in the following Hamiltonian

\begin{equation}
\label{rel.5}
\tilde H = \frac{1}{2m} (\vec p - \frac{e}{c} \vec{A} )^2 - \frac{(e\varphi - D)^2}{2mc^2} + \frac{mc^2}{2}
\end{equation}
\,
\\
describing the {\bf nonrelativistic} motion of the particle of mass $m$ and charge $e$ moving in the electromagnetic field described by the scalar potential $ -\frac{1}{2mc^2e} (e \varphi - D )^{2}$ and vector potential $\vec A$.\\
Due to the condition (\ref{rel.4}) we have to restrict ourselves to the trajectories obeying

\begin{equation}
\label{rel.6}
\frac{1}{2m} (\vec p - \frac{e}{c} \vec A )^2 - \frac{(e\varphi - D)^2}{2mc^2} \leq 0
\end{equation}
\,
\\
By skipping the constant in (\ref{rel.5}) we conclude that the relevant nonrelativistic motion is described by the Hamiltonian

\begin{equation}
\label{rel.7}
\tilde H = \frac{1}{2m} (\vec p - \frac{e}{c}\vec A )^2 - \frac{(e\varphi - D)^2}{2mc^2}
\end{equation}
\,
\\
together with the condition 

\begin{equation}
\label{rel.8}
\tilde H \leq 0.
\end{equation}
\,
\\
Replacing eq. (\ref{rel.5}) by eq. (\ref{rel.7}) leads to the following modification of eq. (\ref{rel.2}):

\begin{equation}
\label{rel.9}
m^{2}\left(\tilde{H}\right)=-\frac{2m\tilde{H}}{c^{2}}
\end{equation}
\,
\\
On the other hand, by virtue of eqs. (\ref{rel.1}), (\ref{rel.3}) and (\ref{rel.9}) one obtains

\begin{equation}
\label{rel.10}
c\sqrt{\left(\vec{p}-\frac{e}{c}\vec{A}\right)^{2}+m^{2}\left(\tilde{H}\right)c^{2}}+e\varphi = D
\end{equation}
\,
\\
As a result we are comparing the motion of \textbf{nonrelativistic} particle of mass $m$, charge $e$ and nonpositive energy $\tilde{E} \leq 0$, moving in the electromagnetic field $\left(-\frac{\left(e \varphi-D\right)^{2}}{2mc^{2}e}, \vec{A}\right)$ with the motion of \textbf{relativistic} particle of the mass $\sqrt{\frac {-2m\tilde{E}}{c^{2}}}$, charge $e$ and energy $D$, moving in the electromagnetic field $\left(\varphi, \vec{A}\right)$. The Hamiltonian describing the relativistic motion reads simply 

\begin{equation}
\label{rel.11}
H=c\sqrt{\left(\vec{p}-\frac{e}{c}\vec{A}\right)^{2}+m^{2}\left(\tilde{E}\right)c^{2}}+e\varphi
\end{equation}
\,
\\
and the trajectories under consideration obey

\begin{equation}
\label{rel.12}
H=D
\end{equation}
\,
\\
As a simple example let us consider the relativistic Coulomb problem (see, for example, \cite{lan}). Choosing $\vec{A}=0 \;\;\; \varphi=e'/r$, $\alpha \equiv ee'$ one finds the relativistic Hamiltonian

\begin{equation}
\label{rel.13}
H=c\sqrt{\vec{p}\,^{2}+m^{2}\left(\tilde{E}\right)c^{2}}+\frac{\alpha}{r}\qquad \left(=D\right)
\end{equation}
\,
\\
and its nonrelativistic counterpart

\begin{equation}
\label{rel.14}
\tilde{H}=\frac{\vec{p}\,^{2}}{2m}-\frac{1}{2mc^{2}}\left(\frac{\alpha}{r}-D\right)^{2}\qquad\left(\leq0\right)
\end{equation}
\,
\\
Due to the spherical symmetry of both Hamiltonians the motions are plane and it is convenient to work in polar coordinates $\left(r,\Theta\right)$. Then one finds 

\begin{equation}
\label{rel.15}
H=c\sqrt{p^{2}_{r}+\frac{p^{2}_{\Theta}}{r^{2}}+m^{2}\left(\tilde{E}\right)c^{2}}\;+\frac{\alpha}{r}\qquad \left(=D\right)
\end{equation}
\,
\\
and

\begin{equation}
\label{rel.16}
\tilde{H}=\frac{p^{2}_{r}}{2m}+\frac{p^{2}_{\Theta}}{2mr^{2}}-\frac{\alpha^{2}}{2mc^{2}r^{2}}+\frac{\alpha D}{mc^{2}r}-\frac{D^{2}}{2mc^{2}}\qquad\left(\leq0\right)
\end{equation}
\,
\\
Again, the constant term on the right hand side of eq. (\ref{rel.16}) may be skipped resulting in the Hamiltonian

\begin{equation}
\label{rel.17}
\tilde{H}=\frac{p^{2}_{r}}{2m}+U_{eff}\left(r\right)\leq\frac{D^{2}}{2mc^{2}}
\end{equation}
\,
\\
\begin{equation}
\label{rel.18}
U_{eff}\left(r\right)\equiv \frac{\alpha D}{mc^{2}r}-\frac{\alpha^{2}}{2mc^{2}r^{2}}+\frac{p^{2}_{\Theta}}{2mr^{2}}=\frac{\alpha D}{mc^{2}r^{2}}+\frac{c^{2}p^{2}_{\Theta}-\alpha^{2}}{2mc^{2}r^{2}}
\end{equation}
\,
\\
\begin{equation}
\label{rel.19}
m^{2}\left(\tilde{E}\right)=-\frac{2m\tilde{E}}{c^{2}}+\frac{D^{2}}{c^{4}}
\end{equation}
\,
\\
The relation between relativistic and nonrelativistic trajectories, described above, can be now confirmed by explicit calculations. For both Hamiltonians the total energy and angular momentum $p_{\Theta}$ are conserved so one can proceed in the standard way by separating variables in energy integral and eliminating time differential $dt$ in favour of angle differential $d\Theta$. For both Hamiltonians one finds the same relation

\begin{equation}
\label{rel.20} 
\frac{dr}{d\Theta}=\pm\frac{\sqrt{2m}\,r^{2}}{\vert p_{\Theta}\vert}\sqrt{E-U_{eff}\left(r\right)}
\end{equation}
\,
\\
provided $E=D$ and eq. (\ref{rel.19}) holds true. This shows explicitly that both sets of trajectories coincide. However, there are some further constraints. If we restrict ourselves to the nonrelativistic motion the only condition for the existence of trajectories follows from eq. (\ref{rel.20}).

\begin{equation}
\label{rel.21} 
E-U_{eff}\left(r\right)\geq 0
\end{equation}
\,
\\
for some $r\in \mathbb{R}_{+}$. On the other hand, the trajectories determined by 
eq. (\ref{rel.20}) can be viewed as describing the relativistic motion if, in addition,

\begin{equation}
\label{rel.22} 
m^{2}\left(\tilde{E}\right)\equiv - \frac{2m\tilde{E}}{c^{2}}+\frac{D^{2}}{c^{4}}>0
\end{equation}
\,
\\
\begin{equation}
\label{rel.23} 
D-\frac{\alpha}{r}=c\sqrt{p^{2}_{r}+\frac{p^{2}_{\Theta}}{r^{2}}+m^{2}\left(\tilde{E}\right)c^{2}}\;\geq 0
\end{equation}
\,
\\
\par In what follows we restrict ourselves to the case of attracting Coulomb potential, $\alpha <0$, leaving the $\alpha > 0$ case to the reader.\\
For  $D\geq 0$ eq. (\ref{rel.23}) is automatically satisfied. Therefore, taking into account the first condition, eq. (\ref{rel.22}), all nonrelativistic trajectories can be viewed as relativistic ones provided the nonrelativistic energy obeys

\begin{equation}
\label{rel.24} 
\tilde{E}\leq \frac{D^{2}}{2mc^{2}}
\end{equation}
\,
\\
Depending on whether $c^{2}p^{2}_{\Theta}-\alpha ^{2}\lessgtr 0$ we find two classes of trajectories: for $c^{2}p^{2}_{\Theta}\leq \alpha^{2}$ the particle falls to the centre,

\begin{equation}
\label{rel.25} 
r=\frac{\alpha^{2}-c^{2}p^{2}_{\Theta}}{c\sqrt{\left(p_{\Theta}D\right)^{2}+m^{2}\left(\tilde{E}\right)c^{2}\left(\alpha^{2}-c^{2}p^{2}_{\Theta}\right)}\;\cosh\;\left(\sqrt{\frac{\alpha^{2}}{\left(cp_{\Theta}\right)^{2}}-1}\;\left(\varphi-\varphi_{0}\right)\right)+\alpha D}
\end{equation}
\,
\\
On the other hand, if $c^{2}p^{2}_{\Theta}>\alpha^{2}$, the relevant trajectory reads

\begin{equation}
\label{rel.26} 
r=\frac{c^{2}p^{2}_{\Theta}-\alpha^{2}}{c\sqrt{\left(p_{\Theta}D\right)^{2}-m^{2}\left(\tilde{E}\right)c^{2}\left(c^{2}p^{2}_{\Theta}-\alpha^{2}\right)}\;\cos\;\left(\sqrt{1-\frac{\alpha^{2}}{\left(cp_{\Theta}\right)^{2}}}\;\left(\varphi-\varphi_{0}\right)\right)+\alpha D}
\end{equation}
\,
\\
The case $D<0$ is more subtle. In particular, if $c^{2}p^{2}_{\Theta}\geq \alpha^{2}$ the nonrelativistic motion is possible for any $\tilde{E}>0$ while no trajectory can be viewed as describing relativistic motion (due to the negativity of kinetic energy). For $\alpha^{2}>c^{2}p^{2}_{\Theta}$ all nonrelativistic energies are allowed, $-\infty<\tilde{E}<\infty$. However, only the trajectories describing the fall to the centre with $\tilde{E}<\frac{D}{2mc}$ can be also viewed as relativistic ones. They are given again by Eq.(\ref{rel.25}) with $D$ now beeing negative.

\section{Darboux and Hietarinta methods}
\par

In the eighties Hietarinta et al. introduced a very nice method, known as coupling-constant metamorphosis, of constructing new Hamiltonians out of the given ones. It consists in exchanging the roles of a coupling constant and energy while preserving the form of unparameterized trajectories in phase space. This method is somehow related to the theorem proven by Darboux in 1889 \cite{dar} (see also \cite{alb}). Both Darboux theorem and Hietarinta method nicely fit into general framework described in Sec. II.

Let us start with the natural Hamiltonian of the form

\begin{equation}
\label{r4.54}
H=\frac{1}{2} g^{ij}(q)p_ip_j - \lambda U(q);
\end{equation}
\,
\\
here $\lambda$ is an arbitrary coupling constant and we follow the nonstandard convention denoting by $-\lambda U$ the potential energy. Adding to the right hand side the constant $\lambda - 1$ and making the replacement $\lambda \rightarrow \tilde H$ we arrive at the new Hamiltonian (cf. eq.  (\ref{r2.3}))

\begin{equation}
\label{r4.55}
\tilde H = \frac{1}{U} \left(\frac{1}{2} g^{ij} p_ip_j - 1\right)
\end{equation}
\,
\\ 
Let

\begin{equation}
\label{r4.56}
q^i=\tilde q\, ^i\left ( \tilde t,\tilde E \right ),\quad p_i=\tilde p_i \left ( \tilde t,\tilde E \right)
\end{equation}
\,
\\
where 

\begin{equation}
\label{r4.59}
\tilde{E}= \frac{1}{U} \left( \frac{1}{2} g^{ij}p_ip_j -1\right) ,
\end{equation}
\,
\\
denote the solutions to the canonical equations for the Hamiltonian (\ref{r4.55}). Then the solutions to the canonical equations for $ H$ can be written in the form (cf. eqs.  (\ref{r2.4}), (\ref{r2.6}), (\ref{r2.7}), (\ref{r2.8}))

\begin{equation}
\label{r4.57}
q^i=q^i\left( t;\lambda =\tilde{E} \right) =\tilde q \,^i \left( \tilde t (t), \tilde E \right) ,\quad p_i=p_i\left( t;\lambda =\tilde{E} \right) =\tilde p_i \left( \tilde t (t), \tilde E \right)
\end{equation}
\,
\\
where
 
\begin{equation}
\label{r4.58}
\odv{\tilde{t}}{t} \equiv \Omega = U.
\end{equation}
\,
\\
The Lagrangians corresponding to $H$ and $\tilde{H}$ read, respectively (up to irrelevant additive constants)

\begin{equation}
\label{r4.60}
L=\frac{1}{2}g_{ij}\dot{q}\,^i\dot{q}\,^j +\lambda U
\end{equation}

\begin{equation}
\label{r4.61}
\tilde{L}=\frac{U}{2}g_{ij} \dot{q}\,^i \dot q \,^j +\frac{1}{U}
\end{equation}
\,
\\
Now, according to the Darboux theorem  \cite{dar}: the unparametrized trajectories in configuration space for the Lagrangians $\tilde L$ (cf. eq.  (\ref{r4.61})) and $L'$ defined by the formula

\begin{equation}
\label{r4.63}
L'= \frac{1}{2} g_{ij}\dot q\,^i \dot q\,^j + U
\end{equation}
\,
\\
coincide and the energies $\tilde E$ and $E'$ corresponding to $\tilde L$ and $L'$, respectively satisfy the relation

\begin{equation}
\label{r4.69}
E'\cdot \tilde{E} = 1.
\end{equation}
\,
\\
Both conclusions follow  from our approach. In fact, the solution ${q'}\,^i = {q'}\,^i(t)$ to the Lagrange equations for $L'$ can be written in terms of the solutions $q^i=q^i(t; \lambda)$ to the Lagrange equations for $L$ as follows

\begin{equation}
\label{r4.64}
q'\, ^i (t) =q^i\left(\frac{1}{\sqrt{\lambda}}t; \lambda\right)
\end{equation}
\,
\\
Putting $\lambda =\tilde E$ and taking into account the first of equations (\ref{r4.57}) we conclude 

\begin{equation}
\label{r4.65}
{q'}\,^i(t) = \tilde q\,^i \left(\tilde t \left( \frac{t}{\sqrt{\tilde E}} \right ), \tilde{E} \right)
\end{equation}
\,
\\
The energy $E'$ corresponding to $L'$

\begin{equation}
\label{r4.68}
E' = \frac{1}{2} g_{ij} \odv{qi'\,^i}{t} \odv{q'\, ^j}{t} - U
\end{equation}
\,
\\
when expressed in term of $\tilde q \,^i$ (cf. eq. (\ref{r4.58})) reads

\begin{equation}
\label{r4.68a}
E'=\frac{U^2}{\tilde E} \frac{1}{2} g_{ij}\odv{\tilde q \,^i}{\tilde t} \odv{\tilde q \,^j}{\tilde t} - U
\end{equation}
\,
\\
By comparing eq. (\ref{r4.68a}) with the energy corresponding to $\tilde L$

\begin{equation}
\label{r4.68b}
\tilde E = \frac{U}{2}g_{ij}\odv{\tilde q \,^i}{\tilde t} \odv{\tilde q \,^j}{\tilde t} - \frac{1}{U}
\end{equation}
\,
\\
we find the relation (\ref{r4.69}). \\
As far as the Hietarinta et al. method is concerned one rewrites the Hamiltonian (1) from Ref.  \cite{hit} by adding the constant

\begin{equation}
\label{r4.70}
H=H_0 - gF +(g-h)
\end{equation}
\,
\\
Letting $g=\tilde{H} \equiv G$ one arrives at the formula (2) from \cite{hit}

\begin{equation}
\label{r4.71}
G=\frac{H_0}{F} - \frac{h}{F}
\end{equation}

\section{Conclusions and outlook}

\par
We proposed a general framework for constructing, for a given Hamiltonian, the whole family of new Hamiltonians sharing the same set of unparameterized trajectories in phase space. Once the symmetry of initial Hamiltonian is known our approach provides the complete characterization of the symmetry structure of new Hamiltonian systems. In fact, eqs. (\ref{r2.11}) and (\ref{r2.14})-(\ref{r2.17}) imply that the integrals of motion for the latter can be immediately constructed provided those corresponding to the initial Hamiltonian are known. 

We have shown that our approach covers many examples considered recently in the literature as well as provides the new ones. We presented only some of them. It is interesting to  note that further, more complicated, examples can be obtained by applying the  algorithm iteratively.

Our procedure provides a far reaching generalization of Hietarinta et al. coupling constant-energy metamorphosis method \cite{hit}. It allows also to give an alternative proof of Darboux theorem.

Moreover, we found an interesting relation between nonrelativistic and relativistic dynamics which, up to our best knowledge, hasn't been discussed so far. As an example, the relativistic motion in attractive Coulomb potential has been considered in some detail; in particular, the family of nonrelativistic potentials yielding the same trajectories was exhibited.

The method presented above is both fairly general and flexible. First, let us note that it can be extended by assuming that the parameters $\lambda_{\alpha}$ depend, apart from the Hamiltonian, on other integrals of motion (when they exist). This opens the way to construct still more general new Hamiltonian systems. It appears also that the judicious choice of the initial Hamiltonian allows to obtain the standard Hamiltonian on locally flat configuration space; however, the global structure of the latter becomes more complicated. 

One can also apply the present approach to the study of geodesics on (pseudo)Riemannian manifolds. In particular, it provides a nice way to construct the Carter integral for null geodesics in Kerr metric. 

What is more, it seems that our procedure can be used to analyze the generalization of Bertrand's theorem, both in the spirit of Perlick's paper \cite{24} and the relativistic motion in central potentials (i.e. when the Hamiltonian is the sum of relativistic kinetic energy and central potential).

Some of these problems will be studied in subsequent publications.

\end{document}